\documentclass[prl, twocolumn, 10pt,a4paper,superscriptaddress,notitlepage, longbibliography]{revtex4-1}

\usepackage{amsmath}
\usepackage{amssymb}
\usepackage{textcomp}
\usepackage{upgreek}
\usepackage{units}
\usepackage{graphicx}
\usepackage{microtype}
\usepackage{enumitem}
\usepackage{xcolor}
\usepackage{gnuplottex}    
\usepackage{multirow}   
\usepackage{physics}
\usepackage{siunitx}
\usepackage{booktabs}
\usepackage{csquotes}

\usepackage{hyperref}
\hypersetup{colorlinks=true, citecolor=blue, urlcolor=blue, linkcolor=blue}

\usepackage{txfonts}

\renewcommand{\vec}[1]{\ensuremath{\boldsymbol{#1}}}

\newcommand{\nk}{{n \vec{k}}}
\newcommand{\mub}{\mu_\mathrm{B}}

\newcommand{\mathpi}{\piup}

\usepackage{pifont}

\begin{document}
\title{
Orbital Magnetic Moment of Magnons
}

\author{Robin R.~Neumann}
\address{Institut f\"ur Physik, Martin-Luther-Universit\"at Halle-Wittenberg, D-06099 Halle (Saale), Germany}
\author{Alexander Mook}
\address{Institut f\"ur Physik, Martin-Luther-Universit\"at Halle-Wittenberg, D-06099 Halle (Saale), Germany}
\affiliation{Department of Physics, University of Basel, Klingelbergstrasse 82, CH-4056 Basel, Switzerland}
\author{J\"{u}rgen Henk}
\address{Institut f\"ur Physik, Martin-Luther-Universit\"at Halle-Wittenberg, D-06099 Halle (Saale), Germany}
\author{Ingrid Mertig}
\address{Institut f\"ur Physik, Martin-Luther-Universit\"at Halle-Wittenberg, D-06099 Halle (Saale), Germany}

\begin{abstract}
    In experiments and applications usually the spin magnetic moment of magnons is considered. In this Paper we identify an additional degree of freedom of magnons: an \emph{orbital} magnetic moment brought about by spin-orbit coupling.
    Our microscopic theory uncovers that spin magnetization $\vec{M}^\mathrm{S}$ and orbital magnetization $\vec{M}^\mathrm{O}$ are independent quantities.
    They are not necessarily collinear; thus, even when the total spin moment is compensated due to antiferromagnetism ($\vec{M}^\mathrm{S} = \vec{0}$), $\vec{M}^\mathrm{O}$ may be nonzero. This scenario of orbital weak ferromagnetism is realized in paradigmatic kagome antiferromagnets with Dzyaloshinskii-Moriya interaction. We demonstrate that magnets exhibiting a magnonic orbital moment are omnipresent and propose transport experiments for probing it. 
\end{abstract}
\date{\today}
\maketitle

% =============================================
%  INTRODUCTION
% =============================================
\paragraph{Introduction.} Textbooks on magnetism introduce spin waves as collective excitations of a magnetically ordered ground state, as epitomized by ferromagnets (Ref.~\onlinecite{Ashcroft1976} among others). The quanta of spin waves -- the magnons -- are typically viewed as local deviations from the ordered state \cite{Bloch1930, Bloch1932}. Within this picture, it appears natural that the magnetic moment carried by magnons has only spatial components that are offered by the ground-state spin texture, because the latter defines the directions relative to which a deviation can occur \cite{Okuma2017}. This implies in particular that collinear magnets feature only magnons whose magnetic moment is along the collinear axis. Likewise the magnetic moments of magnons of coplanar magnets lie within that plane. This reasoning is widely accepted and adopted for a plethora of transport phenomena that involve the magnon magnetic moment, such as the spin Seebeck \cite{Uchida2010}, spin Nernst \cite{Han2016ax, Cheng2016, Kovalev2016, Mook2016c, Shiomi2017, Wang2018anomalous, Mook2018, Mook2019SSESNE, Mook2020MSHE}, and magnon Edelstein effect \cite{Shitade2019Edelstein, Li2019Edelstein} in ferromagnets \cite{Cramer2017} and in both collinear \cite{Seki2015, Wu2016, Lebrun2018} and noncollinear \cite{Mook2017a,Mook2019SSESNE, Flebus2019, Li2020, Mook2020MSHE} antiferromagnets.

In this Paper, we challenge this paradigm by revealing an additional magnonic degree of freedom: their orbital magnetic moment. Overall, the magnetic moment
\begin{align}
    \vec{\mu}_{n,\vec{k}} = -\frac{\partial \varepsilon_{n,\vec{k}}}{\partial \vec{B}} 
    =
    \vec{\mu}_{n,\vec{k}}^\mathrm{S}
    +
    \vec{\mu}_{n,\vec{k}}^\mathrm{O}
    \label{eq:defFullMom}
\end{align}
of a magnon in band $n$ and with momentum $\hbar \vec{k}$ decomposes into two contributions.
These are derived from the explicit and implicit dependence of the magnon energy $\varepsilon_{n,\vec{k}}$ with respect to the magnetic field $\vec{B}$. The first contribution,
\begin{align}
    \vec{\mu}^\text{S}_{n,\vec{k}} \propto - \vec{s}_{n,\vec{k}},
    \label{eq:defSpinMom}
\end{align}
is the spin magnetic moment (SMM) which is proportional to the magnon spin $\vec{s}_{n,\vec{k}}$ \cite{Okuma2017, Li2019Edelstein}. As mentioned above, this is the contribution conventionally referred to as the magnetic moment of magnons.
The second contribution $\vec{\mu}^\text{O}_{n,\vec{k}}$  -- the \emph{orbital} magnetic moment (OMM) -- captures the difference of Eqs.~\eqref{eq:defFullMom} and \eqref{eq:defSpinMom} and is the main object of interest in this Paper.
It is associated with an \emph{implicit} dependence of $\varepsilon_{n,\vec{k}}$ on $\vec{B}$, which arises from the field-dependent relative orientation of the magnetic texture to the structural lattice \footnote{We implicitly assume that the usual Zeeman term is the only field-dependent term in the spin Hamiltonian.} and, hence, requires spin-orbit coupling (SOC). The SMM and the OMM result in macroscopic spin and orbital magnetizations, $\vec{M}^\mathrm{S}$ and $\vec{M}^\mathrm{O}$, respectively. These independent quantities can be  disentangled clearly in the situation of \emph{magnonic orbital weak ferromagnetism}, in which $\vec{M}^\mathrm{S} = \vec{0}$ but $\vec{M}^\mathrm{O} \ne \vec{0}$. Importantly, even if $\vec{M}^\mathrm{O} = \vec{0}$ in equilibrium, the OMM may be addressed by an \emph{orbital Nernst effect of magnons} in nonequilibrium. As a consequence, the complete set of magnonic degrees of freedom may be utilized for insulator spintronics.

% =============================================
%  DEFINITION OF MAGNETIZATION
% =============================================
\paragraph{Identification of the orbital magnetic moment.} We start with a generic spin Hamiltonian
$
	\hat{H} = \hat{H}_\mathrm{spin} + \hat{H}_\mathrm{Zee}
$;
$\hat{H}_\mathrm{spin}$ and $\hat{H}_\mathrm{Zee} = \hbar^{-1} \mu_\mathrm{B} \sum_i \vec{B} \cdot \vec{g}_i \hat{\vec{S}}_i$ describe the spin-spin interactions (where the magnetic field $\vec{B}$ does not enter) and the coupling to the magnetic field (Zeeman term; $\hbar$ reduced Planck constant, $\mu_\mathrm{B}$ Bohr's magneton), respectively. $\vec{g}_i$ is the g-tensor of the spin operator $\hat{\vec{S}}_i$ at site $i$. 
Assuming an ordered ground state with $N$ spins per magnetic unit cell pointing along $\hat{\vec{z}}_n$ ($n=1,\ldots, N$), we perform a truncated Holstein-Primakoff (HP) transformation \cite{Holstein1940} from spin operators to bosonic operators $\hat{a}_i^{(\dagger)}$, yielding $\hat{H} \approx E_0 + \hat{H}_2$. Here, $E_0$ is the classical ground state energy and $\hat{H}_2$ describes noninteracting magnons. After a transformation to magnonic normal modes $\hat{\alpha}_{n,\vec{k}}^{(\dagger)}$ in reciprocal space, we obtain
$
    \hat{H} \approx E_0 + \varDelta E_0 + \sum_{\vec{k}} \sum_{n=1}^N \varepsilon_{n, \vec{k}} \hat{\alpha}^\dagger_{n, \vec{k}} \hat{\alpha}_{n, \vec{k}}
$.
The harmonic zero-point quantum fluctuations, $\varDelta E_0 = \varDelta E_0^{(1)} + \varDelta E_0^{(2)}$, with
$
    \varDelta E_0^{(1)} 
    = 
    - \frac{1}{4} \sum_{\vec{k}} \Tr \boldsymbol{H}_{\vec{k}},
$
and
$
    \varDelta E_0^{(2)} 
    =
    \frac{1}{2} \sum_{\vec{k}} \sum_{n=1}^N \varepsilon_{n, \vec{k}},
$
provide a correction to $E_0$; $\vec{H}_{\vec{k}}$ is the Hamilton matrix. See Supplementary Material (SM) \cite[Sec.~I]{Supplement} for details.

\begin{widetext}
When considering effective spin Hamiltonians, usually the spin magnetization (SM \cite[Sec.~II]{Supplement})
\begin{align}
    \vec{M}^\mathrm{S}(T) 
    = 
    - \frac{\mub}{V} \sum_{\vec{k}} \sum_{n=1}^N \vec{g}_n \hat{\vec{z}}_n \left( S_n - \left\langle \hat{a}^\dagger_{n,\vec{k}}\hat{a}_{n,\vec{k}} \right\rangle \right)
    =
    \underbrace{- \frac{\mub}{V_\text{uc}} \sum_{n=1}^N S_n \vec{g}_n \hat{\vec{z}}_n}_{\vec{M}_0^\mathrm{S}}
    \underbrace{- \frac{\mub}{2 V_\text{uc}} \sum_{n=1}^N \vec{g}_n \hat{\vec{z}}_n}_{\varDelta\vec{M}_0^{\mathrm{S},(1)}}
    \underbrace{+ \frac{1}{2V} \sum_{n=1}^N \sum_{\vec{k}} \vec{\mu}^\mathrm{S}_{n,\vec{k}}}_{\varDelta\vec{M}_0^{\mathrm{S},(2)}}
    \underbrace{+
    \frac{1}{V} \sum_{n=1}^N \sum_{\vec{k}} \vec{\mu}^\mathrm{S}_{n,\vec{k}} \rho(\varepsilon_{n,\vec{k}},T)}_{\vec{M}_2^\mathrm{S}(T)}
    \label{eq:spin_magnetization}
\end{align}
is addressed ($V$ sample volume, $V_\text{uc}$ volume of a unit cell, $\langle \cdot \rangle$ thermodynamic average). $S_n$ and $\vec{g}_n$ are the length and the $g$-tensor of the $n$th spin in the unit cell, respectively. Although $\vec{g}_n$ already incorporates SOC, we denote $\vec{M}^\mathrm{S}$ a ``spin'' magnetization, because the set $\{ \vec{g}_n \}$ merely transforms the directions $\hat{\vec{z}}_n$. The above sum is decomposed into the classical ground state spin magnetization $\vec{M}^\mathrm{S}_0$, its quantum corrections $\varDelta \vec{M}^\mathrm{S}_0 = \varDelta\vec{M}_0^{\mathrm{S},(1)} + \varDelta\vec{M}_0^{\mathrm{S},(2)}$, and into $\vec{M}^\mathrm{S}_2(T)$ which is due to the thermal population of magnons [$\rho(\varepsilon_{n\vec{k}},T) = (\mathrm{e}^{\beta \varepsilon_{n\vec{k}}}-1)^{-1}$ Bose-Einstein distribution function at temperature $T = (k_{\mathrm{B}} \beta)^{-1}$]. Eventually, $\vec{\mu}^\mathrm{S}_{n,\vec{k}}$ is the SMM of magnons in band $n$ with momentum $\vec{k}$ (SM \cite[Sec.~II]{Supplement}).

$\vec{M}^\mathrm{S}(T)$ does not coincide with the thermodynamical definition of magnetization (SM \cite[Sec.~III]{Supplement})
\begin{align}
    \vec{M}(T) 
    =
	- \frac{1}{V} \frac{\partial \varOmega}{\partial \vec{B}}
    =
    \underbrace{-\frac{1}{V} \frac{\partial E_0}{\partial \vec{B}}}_{\vec{M}_0}
    \underbrace{+\frac{1}{4V} \sum_{\vec{k}} \frac{\partial \mathrm{Tr} \boldsymbol{H}_{\vec{k}} }{\partial \vec{B}}}_{\varDelta\vec{M}_0^{(1)} =-\frac{1}{V} \frac{\partial \varDelta E_0^{(1)}}{\partial \vec{B}}}
    \underbrace{+\frac{1}{2V} \sum_{\vec{k}} \sum_{n=1}^N \vec{\mu}_{n, \vec{k}}}_{\varDelta\vec{M}_0^{(2)}=-\frac{1}{V} \frac{\partial \varDelta E_0^{(2)}}{\partial \vec{B}}}
    \underbrace{+\frac{1}{V} \sum_{n=1}^N \sum_{\vec{k}} \vec{\mu}_{n,\vec{k}}  \rho(\varepsilon_{n,\vec{k}},T)}_{\vec{M}_2(T)}
    \label{eq:thermodyanmics}
\end{align}
($\varOmega$ grand potential). $\vec{\mu}_{n, \vec{k}}$ is the full magnonic magnetic moment defined in Eq.~\eqref{eq:defFullMom}. The constituents of $\vec{M}$ are defined in analogy to those of $\vec{M}^\mathrm{S}$.
\end{widetext}

To verify briefly that $\vec{M}^\mathrm{S} \ne \vec{M}$ replace the g-tensor by a scalar ($\vec{g}_n \to g_n$). $\vec{M}^\mathrm{S}$ is then restricted to those spatial components offered by the $\hat{\vec{z}}_n$'s; however, $\vec{M}$ and $\vec{\mu}_{n,\vec{k}}$ are not, because the $\hat{\vec{z}}_n$'s themselves depend on $\vec{B}$. Thus, even if all  $\hat{\vec{z}}_n$'s are collinear (or coplanar), $\vec{\mu}_\nk$ may have an orthogonal component whose integral is nonzero; hence, $\vec{M} \nparallel \vec{M}^\mathrm{S}$.

The observation $\vec{M}_0 = \vec{M}_0^\mathrm{S}$ (SM \cite[Sec.~IV]{Supplement}) allows to trace the difference of $\vec{M}$ and $\vec{M}^\mathrm{S}$ back to the difference between $\vec{\mu}_{n,\vec{k}}$ and $\vec{\mu}_{n,\vec{k}}^\mathrm{S}$.  More precisely, one obtains
$
    \vec{\mu}_{n,\vec{k}} 
    = 
    \vec{\mu}_{n,\vec{k}}^\mathrm{S} 
    + \vec{\mu}_{n,\vec{k}}^\mathrm{O},
$
in which the SMM is derived from the explicit $\vec{B}$ dependence of the Zeeman energy and the OMM (SM \cite[Sec.~V]{Supplement})
\begin{align}
    \vec{\mu}_{n,\vec{k}}^\mathrm{O} 
    = 
    -
    \sum_{m=1}^N
    \sum_{\alpha = x, y, z}
    \frac{\partial \varepsilon_{n,\vec{k}}}{\partial \hat{\vec{\alpha}}_m} \cdot \frac{\partial \hat{\vec{\alpha}}_m}{\partial \vec{B}}
    \label{eq:orbitalmoment}
\end{align}
from the implicit $\vec{B}$ dependence of the local coordinate system $\{\hat{\vec{x}}_n, \hat{\vec{y}}_n, \hat{\vec{z}}_n \}$ \footnote{Field-dependent terms in the spin Hamiltonian beyond the Zeeman term, e.\,g. ring exchange \cite{Zhang2019OrbMagNernst}, yield additional contributions. For a general definition and more details, see SM \cite{Supplement}.}. Such a dependence has to result from SOC (or SOC-like interactions) which couples spins and lattice and therefore motivates the term ``orbital'' moment. The orbital magnetization 
\begin{align}
    \vec{M}^\mathrm{O}(T)
    = 
    \varDelta \vec{M}_0^{\mathrm{O},(1)} + 
    \underbrace{\frac{1}{2V} \sum_{\vec{k}} \sum_{n=1}^N \vec{\mu}_{n, \vec{k}}^\mathrm{O} }_{\varDelta \vec{M}^{\mathrm{O},(2)}_0}
    \underbrace{+
    \frac{1}{V} \sum_{n=1}^N \sum_{\vec{k}}  \vec{\mu}_{n, \vec{k}}^\mathrm{O} \rho(\varepsilon_{n\vec{k}},T)}_{\vec{M}_2^\mathrm{O}(T)}
    \label{eq:orbitalmag}
\end{align}
is absent in the classical ground state, since it is exclusively due to quantum ($\varDelta \vec{M}_0^{\mathrm{O},(1)} + \varDelta \vec{M}^{\mathrm{O},(2)}_0$) and thermal fluctuations ($\vec{M}_2^\mathrm{O}(T)$).

In what follows, we assume scalar $g$-factors and include SOC exclusively via spin-spin interactions.

% =============================================
%  KAGOME
% =============================================

\begin{figure}
    \includegraphics[width = 1.0\columnwidth]{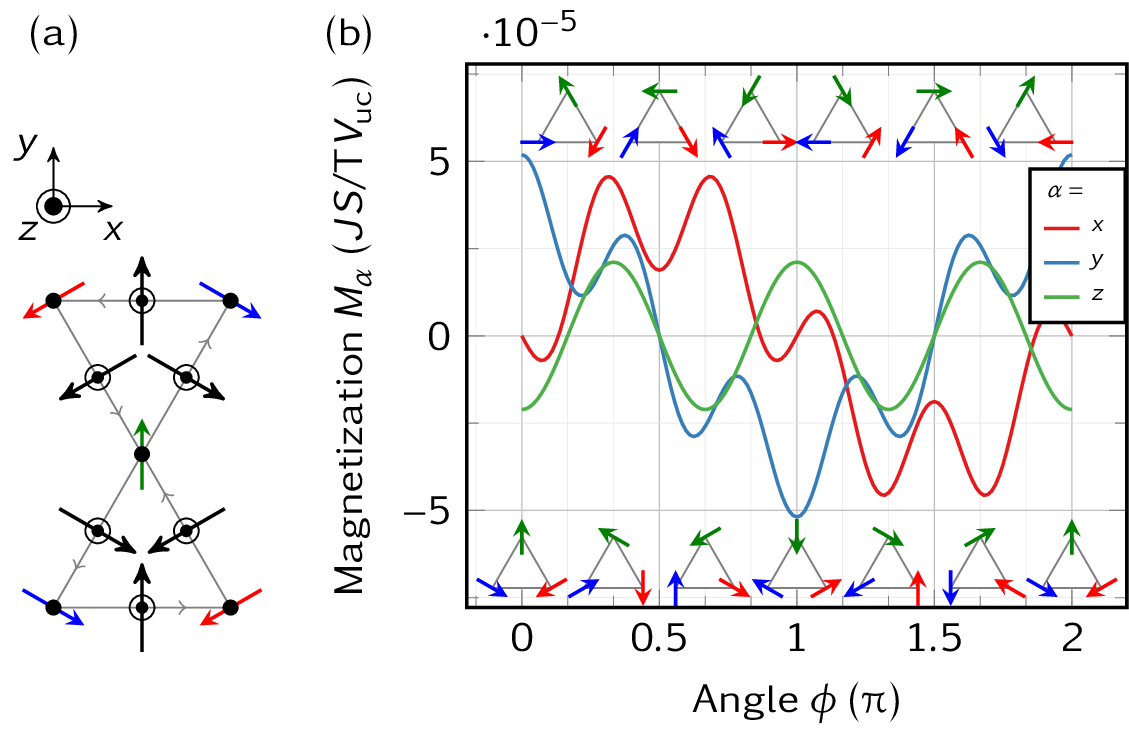}
    \caption{Weak ferromagnetism due to quantum fluctuations in the NVC phase on the kagome lattice. (a) Structural lattice with $\hat{\vec{z}}_n$ ($n = 1, 2, 3$) indicated by colored arrows and DMI vectors by black arrows. 
    (b) Magnetization components $M_\alpha$ ($\alpha = x, y, z$) at zero temperature in dependence on $\phi$. Since $|\vec{M}| \ne 0$, quantum weak ferromagnetism is omnipresent; the out-of-plane magnetization $M_z = M_z^\mathrm{O}$ is attributed to an orbital moment. Parameters read $J = \SI{3.18}{\milli\electronvolt}$, $S = 5/2$, $D_z = 0.062 J$, $D_\parallel = \SI{1}{\milli\electronvolt}$.}
    \label{fig:kagome}
\end{figure}

\paragraph{Orbital magnetic moments in equilibrium.} First, we demonstrate how OMM can be probed in equilibrium as a contribution to weak ferromagnetism. This phenomenon is usually described at the level of classically antiferromagnetic spin textures that exhibit a small canting, e.\,g., due to Dzyaloshinskii-Moryia interaction (DMI) \cite{Dzyaloshinsky58, Moriya60} ($\vec{M}^\mathrm{S}_0 \ne \vec{0}$). Here, we predict pure \emph{orbital} weak ferromagnetism: $\vec{M}^\mathrm{S}_0 = \vec{0}$ but $\vec{M}^\mathrm{O} \ne \vec{0}$. A system of choice is a kagome antiferromagnet [Fig.~\ref{fig:kagome}(a)] with the spin Hamiltonian
\begin{align}
    \hat{H} = 
     \frac{1}{2 \hbar^2} \sum_{\langle ij \rangle} \left( - J \hat{\vec{S}}_i \cdot \hat{\vec{S}}_j
    + \vec{D}_{ij} \cdot \hat{\vec{S}}_i \times \hat{\vec{S}}_j \right)
    + \frac{g \mu_\mathrm{B}}{\hbar} \vec{B} \cdot \sum_i \hat{\vec{S}}_i, \label{eq:spinHam-kagome}
\end{align} 
whose classical phase diagram was derived in Ref.~\onlinecite{Elhajal2002}.  Each spin interacts with its four neighbours via antiferromagnetic exchange $J < 0$ and SOC-induced DMI. The DMI vectors $\vec{D}_{ij}$ are orthogonal to the respective bond [black arrows in Fig~\ref{fig:kagome}(a)] and have both an in-plane ($D_\parallel$) and an out-of-plane components ($D_z$).  For $D_z > 0$ and $|D_\parallel|$ below a critical value, the classical magnetic ground state is an antiferromagnetic \emph{coplanar} texture with negative vector chirality (NVC) \cite{Elhajal2002} [colored arrows in Fig.~\ref{fig:kagome}(a)]. The classical spin magnetization vanishes ($\vec{M}^\mathrm{S}_0 = \vec{0}$). For $E_0$ exhibits an accidental degeneracy under global in-plane rotation of all spins, we perform an order-by-disorder study with respect to the rotation angle $\phi$ [insets in Fig.~\ref{fig:kagome}(b)]. Both quantum and thermal fluctuations select the $\phi = 0$ texture [Fig.~\ref{fig:kagome}(a)] and its $\mathpi/3$ rotations over any other rotated texture (SM \cite[Sec.~VI]{Supplement}). Nonetheless, we proceed with studying all textures.

\begin{table}
    \centering
    \caption{Magnetic point group and symmetry-imposed shape of $\vec{M}$ for NVC phases with $\phi = 0$ and $\phi = \mathpi/2$.}
    \begin{tabular}{ccc}
        \toprule
        Angle $\phi$ & $0$ & $\mathpi/2$ \\
        \midrule
        Magnetic point group & $2'/m'$ & $2/m$\\
        Compatible magnetization & $\mqty(0 & M_y & M_z)$ & $\mqty(M_x & 0 & 0)$
        \\
        \bottomrule
    \end{tabular}
    \label{tab:symmetry_kagome}
\end{table}

\begin{figure}
    \centering
    \includegraphics[width=\columnwidth]{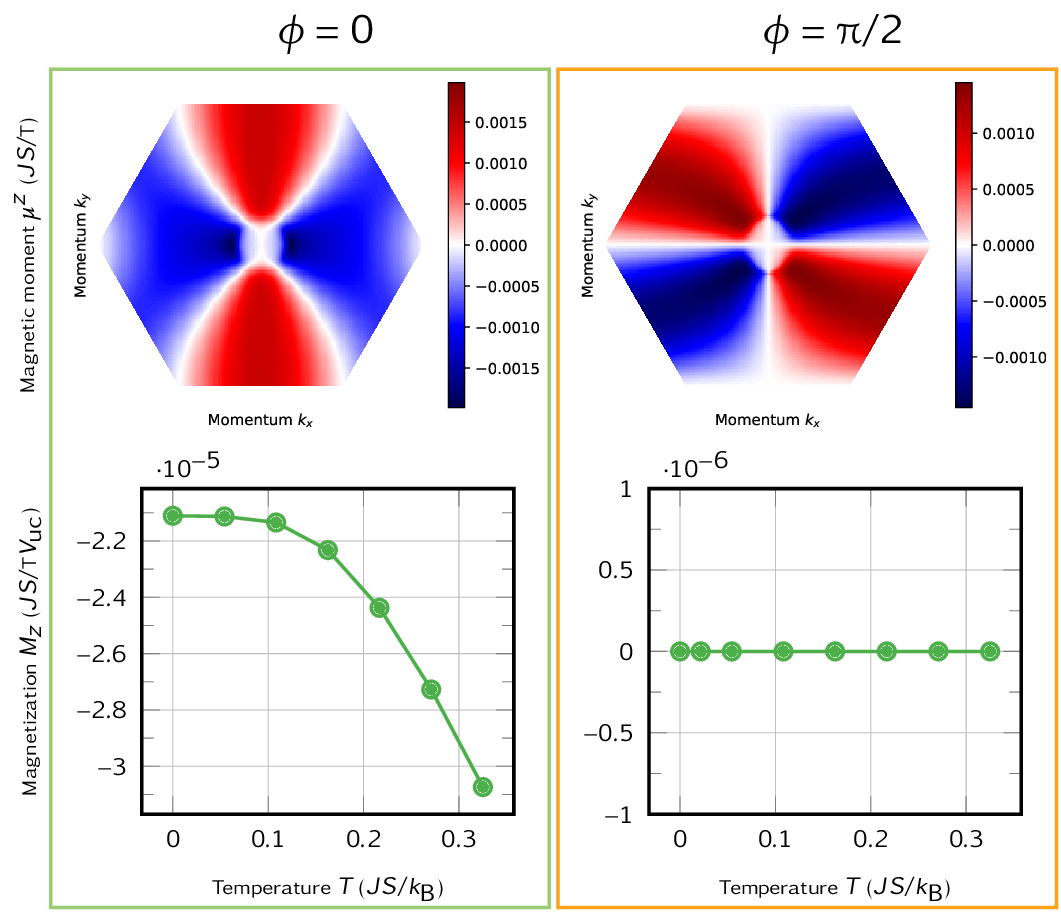}
    \caption{Top: momentum-dependent OMM $\mu_{1,\vec{k},z}$ of the lowest magnon band ($n = 1$) in the Brillouin zone of the kagome antiferromagnet in the NVC phase for $\phi = 0$ (left) and $\phi = \mathpi / 2$ (right). Bottom: temperature dependence of the orbital magnetization $M_z$. Parameters as in Fig.~\ref{fig:kagome}.
    }
    \label{fig:kagome_selected_phases}
\end{figure}

For the discussion we single out the phases for $\phi = 0$ with magnetic point group $2'/m'$ (the prime indicates additional time reversal) and $\phi = \mathpi/2$ with $2/m$ \cite{Mook2019}. In both cases the two-fold rotation axis is along the $x$ direction and the mirror plane coincides with the $yz$ plane. Both groups are compatible with ferromagnetism (Tab.~\ref{tab:symmetry_kagome}). Besides an in-plane magnetization, the $\phi = 0$ phase is also compatible with a nonzero $M_z$. Since $\hat{z}_{n,z} = 0$ by construction, any nonzero $M_z = M_z^\mathrm{O}$ must be attributed to an orbital moment.
% (Arbitrary $\phi$ allows for any magnetization component.)

This symmetry analysis is fully confirmed by the magnetization  calculated from Eq.~\eqref{eq:thermodyanmics} [Fig.~\ref{fig:kagome}(b)]. Although $\vec{M}_0(\phi) = \vec{M}^\mathrm{S}_0(\phi) = \vec{0}$ for all $\phi$, the quantum-corrected magnetization is \emph{never} compensated: $|\varDelta \vec{M}_0(\phi)| \ne 0$. Hence, the quantum fluctuations cause the weak ferromagnetism, of both \emph{spin and orbital} origin for $M_x$ and $M_y$ but of \emph{pure orbital} origin for $M_z$. This finding complements classical analyses of kagome antiferromagnets \cite{Elhajal2002} and shows that even the NVC phase exhibits weak ferromagnetism without the need of higher-order anisotropies beyond DMI \footnote{For parameters of CdCu$_3$(OH)$_6$(NO$_3$)$_2\cdot$H$_2$O---we used $S=1/2$, $g=2.1$, $D_\parallel = D_z = 0.1J$, and $J=\unit[3.87]{meV}$ and $\phi=\mathpi/2$---we find that the weak moment accounts for $\lesssim 1\,\%$ of the experimentally found $\unit[7.93\times10^{-2}]{\mub/Cu}$
\cite{Okuma2017InverseChiral}. Similar estimates may apply to Ca-kapellasite \cite{Ihara2020} and YCu$_3$(OH)$_6$Cl$_3$ \cite{Zorko2019}.}. It is also a counterexample to the common belief that quantum fluctuations only reduce the magnitude of the ordered moment.

The microscopic origin of $M_z \ne 0$ can be studied on the basis of the OMM $\mu_{1,\vec{k},z} = \mu_{1,\vec{k},z}^\mathrm{O}$ of the lowest magnon band ($n = 1$) for both phases (top row of Fig.~\ref{fig:kagome_selected_phases}; recall $\mu_{1,\vec{k},z}^\mathrm{S} =0$). Already an ``ocular integration'' over the Brillouin zone reveals that $M_{z}(T) = M_{z}^\mathrm{O}(T)$ from Eq.~\eqref{eq:orbitalmag} must be either nonzero ($\phi = 0$) or zero ($\phi = \mathpi / 2$), an observation confirmed by numerical integration (bottom row of Fig.~\ref{fig:kagome_selected_phases}). For the $\phi = 0$ phase  $|M_z(T)|$ increases in absolute value with temperature, showing that thermal fluctuations enhance the quantum mechanical weak moment (the $T$ dependence of $M_x$ and $M_y$ is detailed in SM \cite[Sec.~VI]{Supplement}).

That SOC is causing the orbital moment is supported by noting that $\mu_{n,\vec{k},z}, M_{z}(T) \to 0$ as $D_\parallel \to 0$ (not shown). If $D_\parallel = 0$ the kagome plane is an $m'$ plane, which renders $M_z$ zero by symmetry. Hence, in the absence of SOC-induced spin-spin interactions, the orbital magnetization vanishes.

% =============================================
%  PYROCHLORE
% =============================================
\paragraph{Orbital magnetic moments in nonequilibrium.} Having established signatures of OMMs at equilibrium, we now focus on nonequilibrium and consider as an example transport of magnetic moment -- rather than spin -- in the pyrochlore ferromagnet Lu$_2$V$_2$O$_7$. The spin Hamiltonian \cite{Onose2010}
\begin{align}
    \hat{H} = 
     \frac{1}{2 \hbar^2} \sum_{\langle ij \rangle} \left( - J \hat{\vec{S}}_i \cdot \hat{\vec{S}}_j
    + \vec{D}_{ij} \cdot \hat{\vec{S}}_i \times \hat{\vec{S}}_j \right)
    + \frac{g \mu_\mathrm{B}}{\hbar} \vec{B} \cdot \sum_i \hat{\vec{S}}_i, \label{eq:spinHam-pyro}
\end{align}
includes DMI vectors $\vec{D}_{ij} = D \hat{\vec{n}}_{ij} \times \hat{\vec{e}}_{ij}$ that are perpendicular to both the bonds $\hat{\vec{e}}_{ij}$ and the normal $\hat{\vec{n}}_{ij}$ of the cube that surrounds that tetrahedron the bond belongs to \cite{Elhajal2005}. For $J > 0$, collinear ferromagnetism is found, $\hat{\vec{z}}_n = -\hat{\vec{b}} = -\vec{B} / B$ ($n = 1,\ldots,4$), and quantum fluctuations are absent, $\varDelta \vec{M}_0 = \vec{0}$.

\begin{figure}
    \centering
    \includegraphics[width=1\columnwidth]{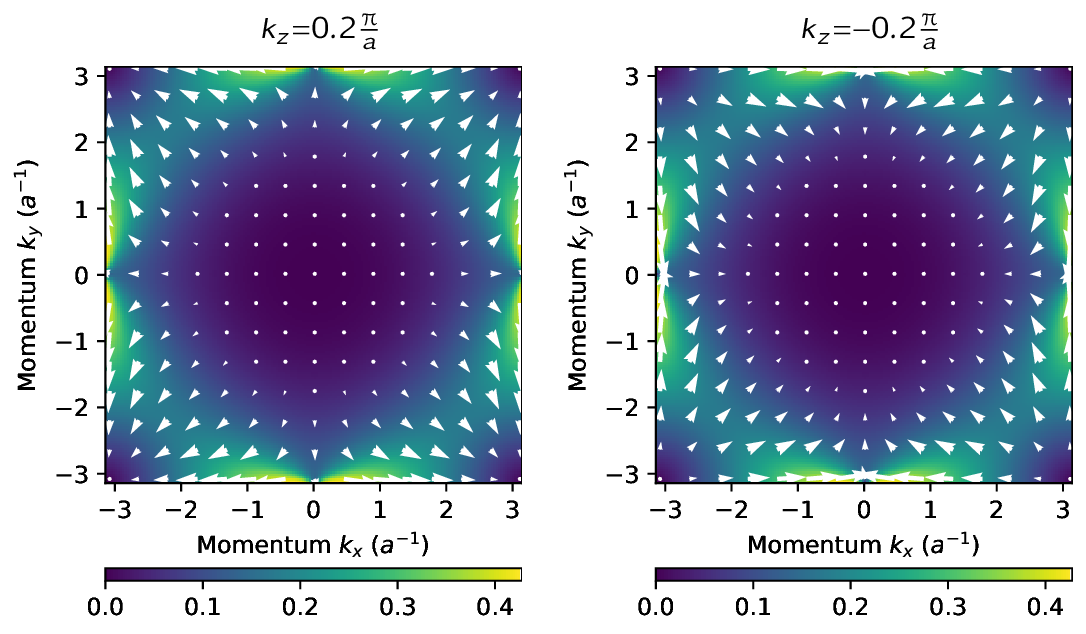}
    \caption{Orbital magnetic moments $\mu_{\vec{k},x}$ and $\mu_{\vec{k},y}$ of the lowest magnon band of the pyrochlore ferromagnet Lu$_2$V$_2$O$_7$. The color scale represents $\sqrt{\mu_{\vec{k},x}^2 + \mu_{\vec{k},y}^2}$ (in units of $\mu_\text{B}$) in two selected $k_x$-$k_y$ planes: $k_z = 0.2\mathpi/a$ (left) and $k_z = -0.2\mathpi/a$ (right); $a$ lattice constant.}
    \label{fig:PyroMagMom}
\end{figure}

The application of a magnetic field $\vec{B} = (0,0,B_z)$ \footnote{This field models an anisotropy; see SM \cite[Sec.~VII]{Supplement} for details.} results in $\mu_{n,\vec{k},z} = \mu_{n,\vec{k},z}^\mathrm{S} = g \mub$ and $\mu_{n,\vec{k},\alpha} = \mu_{n,\vec{k},\alpha}^\mathrm{O} = O(D)$ for $\alpha = x, y$. Hence, the constant $z$ component of $\vec{\mu}_{n,\vec{k}}$ is a SMM\@. The $x$ and $y$ components are OMMs, however, which for positive (negative) $k_z$ resembles a sink-like (source-like) vector field, as depicted in Fig.~\ref{fig:PyroMagMom}.

\begin{figure*}
    \centering
    \includegraphics[width=\textwidth]{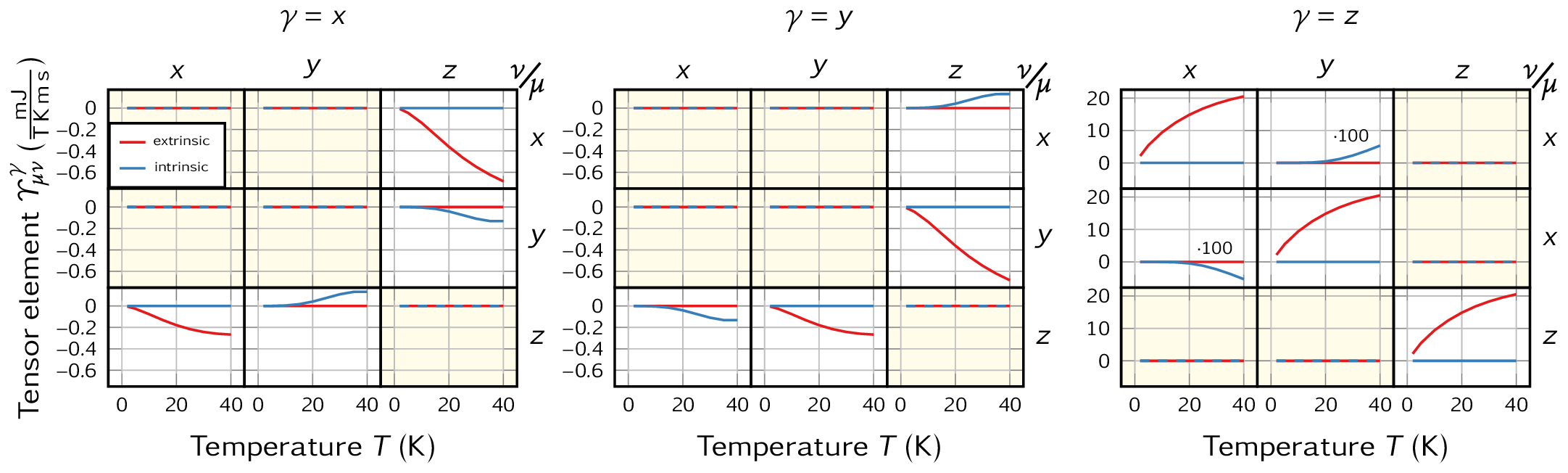}
    \caption{Transport of magnetic moment in the pyrochlore ferromagnet Lu$_2$V$_2$O$_7$. The temperature dependence of all 27 elements of the response tensors $\varUpsilon^\gamma$ are depicted: $\gamma = x$ (left), $\gamma = y$ (center), and $\gamma = z$ (right). In each of the 3-by-3 subfigures, rows (columns) represent the current direction $\mu$ (direction $\nu$ of the temperature gradient). The yellow background highlights vanishing elements. For the extrinsic contributions (red lines), a transport relaxation time $\tau_{n \vec{k}} = \hbar / (\alpha \varepsilon_{n \vec{k}})$ with $\alpha = \num{0.05}$ is assumed. Intrinsic contributions (blue lines) are calculated in the so-called clean limit. Parameters read $J = \SI{-7.99}{\milli\electronvolt}$, $S = 1/2$, $D = \SI{0.5659}{\milli\electronvolt}$, $a = \SI{2.49}{\angstrom}$ \cite{Riedl2016, Onose2010}, and $B_z = \SI{-0.69}{\tesla}$. The ordering temperature of Lu$_2$V$_2$O$_7$ is $\unit[70]{K}$ \cite{Onose2010}.}
    \label{fig:pyrochlor}
\end{figure*}

In equilibrium, the OMM integrates to zero, $\vec{M}_2^\mathrm{O}(T) = \vec{0}$.  However, in \emph{nonequilibrium}, it is transported in transverse direction to a temperature gradient $\vec{\nabla} T$. In other words, this is a Nernst effect (NE) for magnetic moment rather than for spin. Its analysis focuses on the response tensor $\varUpsilon^\gamma$ which relates the nonequilibrium current density of the magnetization with the temperature gradient:
$
    \langle j^\gamma_\alpha \rangle = \varUpsilon^\gamma_{\alpha \beta} ( -\nabla_\beta T )
$
with $\alpha, \beta, \gamma = x, y ,z$.

\begin{table}
    \centering
    \caption{Shape of response tensors $\varUpsilon^\gamma$ ($\gamma = x, y, z$) for the magnetic point group $4/mm'm'$. A subscript ``e'' (``o'') indicates elements that are even (odd) under magnetization reversal.}
    \begin{tabular}{ccc}
        \toprule
        $\varUpsilon^x$ & $\varUpsilon^y$ & $\varUpsilon^z$\\
        \midrule
        $\begin{pmatrix}
            0 & 0 & \varUpsilon_\mathrm{o} \\
            0 & 0 & -\varUpsilon_\mathrm{e} \\
            \varUpsilon'_\mathrm{o} & -\varUpsilon'_\mathrm{e} & 0
        \end{pmatrix}$ &
        $\begin{pmatrix}
            0 & 0 & \varUpsilon_\mathrm{e} \\
            0 & 0 & \varUpsilon_\mathrm{o} \\
            \varUpsilon'_\mathrm{e} & \varUpsilon'_\mathrm{o} & 0
        \end{pmatrix}$ &
        $\begin{pmatrix}
            \tilde{\varUpsilon}_\mathrm{o} & \tilde{\varUpsilon}_\mathrm{e} & 0 \\
            -\tilde{\varUpsilon}_\mathrm{e} & \tilde{\varUpsilon}_\mathrm{o} & 0 \\
            0 & 0 & \tilde{\varUpsilon}'_\mathrm{o}
        \end{pmatrix}$ \\
        \bottomrule
    \end{tabular}
    \label{tab:symmetry_pyrochlor}
\end{table}

Pyrochlore ferromagnets magnetized in $z$ direction belong to the magnetic point group $4/mm'm'$, which dictates the shape of  $\varUpsilon^\gamma$ (Tab.~\ref{tab:symmetry_pyrochlor}).
Tensor elements that are even upon magnetization reversal (subscript ``e'') are associated with intrinsic contributions to the transport, whereas odd elements (subscript ``o'') are associated with extrinsic contributions \cite{Zelezny2017, Mook2019SSESNE}. 

The elements $\tilde{\varUpsilon}_\mathrm{o}$ and $\tilde{\varUpsilon}'_\mathrm{o}$ of $\varUpsilon^z$ comprise a spin Seebeck effect, while $\tilde{\varUpsilon}_\mathrm{e}$ indicates an anomalous spin Nernst effect (SNE) which is associated with spin-polarized transverse particle currents caused by the Berry curvature \cite{Katsura2010, Onose2010, Matsumoto2011, Matsumoto2011a, Matsumoto2014, Mook14a, Mook2018, Mook2019}.

Besides transport of the $z$-component, symmetry admits transport of $x$- and $y$-components as well ($\varUpsilon^x$ and $\varUpsilon^y$ in Tab.~\ref{tab:symmetry_pyrochlor}).
$\varUpsilon_\mathrm{e}$ ($\varUpsilon_\mathrm{e}'$) comprises an anomalous SNE with mutual orthogonality of force, current, and moment directions, whereas $\varUpsilon_\mathrm{o}$ ($\varUpsilon_\mathrm{o}'$) indicates a magnetic SNE \cite{Mook2020MSHE}. Since the $x$- and $y$-components are OMMs, the respective SNEs could be termed ``magnonic \emph{orbital} Nernst effects''.

The above symmetry analysis suggests straightaway an experimental setup for probing OMMs. In a finite pyrochlore sample with $-\vec{\nabla} T \parallel \vec{M} \parallel \vec{z}$, OMM accumulates at the surfaces parallel to $\vec{M}$ ($xz$ and $yz$ surfaces). The resulting surface-located nonequilibrium tilt on $\vec{M}$, conceivably measured by magnetooptical Kerr microscopy, would clearly indicate transport of magnonic orbital magnetization.

We support the above analysis by calculating numerically all 27 elements of $\varUpsilon^\gamma$ within Kubo transport theory (Fig.~\ref{fig:pyrochlor}; SM \cite[Sec.~VII]{Supplement}). Vanishing elements (marked by yellow background) agree with the zeroes in Tab.~\ref{tab:symmetry_pyrochlor}; and so does the either intrinsic (blue, ``e'') or extrinsic (red, ``o'') character. Except for the diagonal elements of $\varUpsilon^z$, all elements scale with the strength $D$ of the DMI, because DMI causes either a nonzero Berry curvature ($\tilde{\varUpsilon}_\mathrm{e}$) or OMMs ($\varUpsilon_\mathrm{e}$, $\varUpsilon'_\mathrm{e}$, $\varUpsilon_\mathrm{o}$, $\varUpsilon'_\mathrm{o}$). With an orbital Nernst conductivity $\varUpsilon^x_{xz} \approx \unit[-0.4]{mJ/(TKms)}$ at $T=\unit[20]{K}$ and $\nabla_z T = \unit[25]{K/mm}$, we find $\langle j^x_{x} \rangle \approx \unit[10]{J/(Tm^2s)}$ (in units of spin, this corresponds to $\hbar \langle j^x_{x} \rangle / \mub \sim \unit[10^{-10}]{J/m^2}$).

% =============================================
%  CHIRAL and DIPOLAR MAGNETS
% =============================================
Another class of magnets lends itself support for nontrivial magnonic OMMs: chiral magnets, like Cu$_2$OSeO$_3$, which hold a prominent place in skyrmion research \cite{Seki2012}.  Their DMI-induced magnonic OMM $\vec{\mu}^\mathrm{O}_{\vec{k}}$ is nonzero, but integrates to $\vec{M}^\mathrm{O} = \vec{0}$ in equilibrium. Due to broken centrosymmetry, however, a magnon current caused by $-\vec{\nabla} T$ exerts a torque on $\vec{M}$ \cite{Manchon2014, Kovalev2016}, an effect that can be explained as an orbital version of the magnon Edelstein effect proposed in Ref.~\onlinecite{Li2019Edelstein}; see SM \cite[Sec.~VIII]{Supplement}.

Dipolar interactions couple spins to the lattice as well. A magnonic OMM -- or better: dipolar magnetic moment -- could be identified as follows. Magnons with $\vec{k} \nparallel \vec{M}$ carry nonzero $\vec{\mu}^\mathrm{O}_{\vec{k}} \perp \vec{M}$. Again, $\vec{M}^\mathrm{O} = \vec{0}$ in equilibrium, but a dipolar-driven ``orbital'' Nernst effect should show up for symmetry reasons, for example in yttrium iron garnet (YIG); see SM \cite[Sec.~IX]{Supplement}.

% =============================================
%  DISCUSSION AND CONCLUSION
% =============================================
\paragraph{Synopsis.} We introduced the orbital magnetic moment of magnons and proposed two experimental signatures: (\i) weak ferromagnetic orbital moment in equilibrium and (\i\i)  accumulation of orbital magnetic moment in nonequilibrium due to a magnonic orbital Nernst effect. Since the latter has the same symmetry as the spin Hall effect \cite{Sinova2015}, it should occur in \emph{any} magnet with large enough SOC or dipolar interactions. Hence, our results pave a way for an all-insulator magnonic spin-orbit torque.

% =============================================
%  ACKNOWLEDGMENTS
% =============================================

\begin{acknowledgments}
\paragraph{Acknowledegments.}
This work is supported by CRC/TRR $227$ of Deutsche Forschungsgemeinschaft (DFG).

\paragraph{Note added.}
The magnonic OMM defined as the difference between total moment and SMM applies to any spin Hamiltonian. For the Hamiltonians discussed in this work, it can be traced back to the dependence of the local coordinate axes on the magnetic field as written in Eq.~\ref{eq:orbitalmoment}. The ``topological orbital moment'' and the resulting orbital Nernst effect of magnons discussed in Ref.~\onlinecite{Zhang2019OrbMagNernst}, both of which rely on a special type of spin interaction, namely three-spin ring exchange, is also captured by Eq.~\ref{eq:defFullMom} and would appear as an additional contribution in Eq.~\ref{eq:orbitalmoment}. However, in the frame of this work, we confined ourselves to bilinear spin-spin interactions.
\end{acknowledgments}

% =============================================
%  BIBLIOGRAPHY
% =============================================

\bibliography{short,newrefs}

\end{document}